\documentclass[prb, twocolumn,showpacs,superscriptaddress]{revtex4-1}
\usepackage[slovak, english]{babel}
\usepackage[cp1250]{inputenc}
\usepackage{graphicx}
\usepackage{dcolumn}
\usepackage{bm}
\usepackage{amssymb}
\usepackage{amsmath}
\usepackage{color}
\usepackage{verbatim}

\begin{document}
\title{{YbPd$_2$In: a promising candidate to strong entropy
accumulation at very low temperature}}

\author{F. Gastaldo} \affiliation{Department of Chemistry,
University of Genova, Via Dodecaneso 31, Genova, Italy}
\author{S. Gabani}
\affiliation{Institute of Experimental Physics, SAS, Ko\v{s}ice, Slovakia}
\author{A. D\v{z}ubinsk\'a}
\affiliation{Faculty of Natural Sciences, P.J. \v{S}af\'arik University, Ko\v{s}ice, Slovakia}
\author{M. Reiffers}
\affiliation{Faculty of Humanities and Natural Sciences, University of PreĹˇov, 17. novembra 1, PreĹˇov, Slovakia}
\author{G. Prist\'a\v{s}}
\affiliation{Institute of Experimental Physics, SAS, Kosice, Slovakia}
\author{I. \v{C}url\'ik}
\affiliation{Faculty of Humanities and Natural Sciences, University of PreĹˇov, 17. novembra 1, PreĹˇov, Slovakia}
\author{P. Skyba}
\affiliation{Institute of Experimental Physics, SAS, Ko\v{s}ice, Slovakia}
\author{M. Clovecko}
\affiliation{Institute of Experimental Physics, SAS, Ko\v{s}ice, Slovakia}
\author{F. Vavrek}
\affiliation{Institute of Experimental Physics, SAS, Ko\v{s}ice, Slovakia}
\author{J. G. Sereni}
\affiliation{Low Temperature Division, CAB-CNEA, 8400 San Carlos de Bariloche, Argentina}
\author{M. Giovannini}
\affiliation{Department of Physics and CNR-SPIN, University of Genova, Via Dodecaneso 33, Genova, Italy}

\begin{abstract}

We report on synthesis, crystal structure, magnetic, thermodynamic
and transport properties of the new compound YbPd$_2$In,
crystallizing as a Heusler structure type. A trivalent state of
the rare earth was determined by fitting the magnetic
susceptibility with a Curie-Weiss law. This compound is
characterized by showing very weak magnetic interactions and a
negligible Kondo effect. A specific heat jump was observed at
$T\approx 250$\,mK, followed at higher temperature by a power law
decrease of $C_P(T)/T$. The resulting large electronic entropy
increase at very low temperature is rapidly shifted to higher
temperature by the application of magnetic field. This magnetocaloric
effect places YbPd$_2$In as a very good candidate for adiabatic
demagnetization cooling processes.

\end{abstract}
\pacs{$71.27.+a; 75.30.-m; 75.30.Mb$}
\maketitle

\section{Introduction}

Along decades, cerium and ytterbium intermetallics attracted
continuous attention owing to the variety of anomalous physical
phenomena discovered on their compounds~\cite{16Ste, 03Car, 16Wu, 09Car}.
Recently notable examples of unique properties have been found in
the Yb-T-X systems (T $=$ transition metals, X $=$ p-type
elements). For instance, in a systematic search for new
ytterbium-palladium indides and stannides ~\cite{10Gio, 17Gas}, we
have synthesized Yb$_2$Pd$_2$Sn, where two quantum critical points
(QCPs) occur under pressure ~\cite{11Mur, 17Yam} and under Sn/In
doping ~\cite{05Bau}.

  Prominent examples with platinum are given by the hexagonal
YbPt$_2$Sn and the cubic Heusler YbPt$_2$In ~\cite{14Gru}.
Interestingly, although crystal structures are different, these
two compounds exhibit many common features, like stable trivalent
Yb$^{3+}$ magnetic moments in a framework of neglibible Kondo
effect and very weak exchange interactions. Moreover, both
compounds are characterized by similar trends of  $C_m/T$ by
temperature decreasing: an increase of $C_m/T$ according of a
power-law temperature dependence below 2 K, followed by broad
anomalies at around 200 mK. It is worth noting that in these
compounds $C_m/T$ reaches record values up to 14 J/mol K$^2$.
Similar features were found in YbCu$_{5-x}$Au$_x$ ($0.4<x<0.7$),
YbCo$_2$Zn$_{20}$ and YbBiPt, but showing a plateaux in $C_m/T(T
\to {0})$ at ~$\approx 7$ J/mol K$^2$ ~\cite{14Cur, 15Gio, 17Ser},
all implying high values of entropy increase at very low
temperature.

  The current interpretation for these unusual behaviors is that
long range magnetic order is inhibited by very weak magnetic
exchange or by magnetic frustration of Yb atoms placed in 2D
(triangular) or 3D (tetrahedra) networks. Coincidentally,
only very low Kondo effect affects their robust magnetic moments.

In some of these compounds, e.g. YbPt$_2$Sn and YbCo$_2$Zn$_{20}$,
it was shown that the low temperature magnetic entropy is
strongly shifted to higher temperature by applying magnetic
fields, offering the interesting perspective to use these
materials as efficient metallic refrigerant for adiabatic
magnetization cooling ~\cite{17Yam, 16Tok}.

In this work we report experimental results on crystal structure
and physical properties of the new cubic Heusler indide YbPd$_2$In
which, based on the present results of this paper, behaves similarly 
to the other members of the YbT$_2$X (T = Pt or Pd and X = Sn or In) family of compounds.

\section{Experimental details}

YbPd$_2$In polycrystalline samples, each with a total weight of
1.2 g, have been prepared by weighting the stoichiometric amount
of elements with the following nominal purity: Yb $99.993$ mass
\% (pieces, Yb/TREM purity, Smart Elements GmbH, Vienna, Austria),
Pd 99.5 mass \% (foil, Chimet, Arezzo, Italy), In 99.999
mass \% (bar). In order to avoid the loss of ytterbium during the
melting because of their high vapor pressure, the proper amounts
of pure elements were enclosed in small tantalum crucibles sealed
by arc welding under pure argon atmosphere. The samples were
synthesized in an induction furnace under a stream of pure argon
and annealed in a resistance furnace at 650 °C for three weeks.
Finally the samples were quenched in cold water and characterized
by optical and scanning electron microscopy (SEM) (EVO 40, Carl
Zeiss, Cambridge, England), equipped with an electron probe
microanalysis system based on energy dispersive X–ray spectroscopy
(EPMA – EDXS). For the quantitative and qualitative analysis an
acceleration voltage of 20 keV for 100 s was applied, and a cobalt
standard was used for calibration. The X-ray intensities were
corrected for ZAF effects. The annealed samples were crushed,
powdered under pure acetone inside an agate mortar and studied by
powder X-ray diffraction (XRD). The XRD data were collected at
room temperature using the X’Pert MPD diffractometer (Philips,
Almelo, The Netherlands) equipped with a graphite monochromator
installed in the diffracted beam (Bragg Brentano, CuK$\alpha$
radiation). The theoretical powder pattern was calculated with 
the Powder-Cell program ~\cite{96kra}. The FULLPROF
program ~\cite{93rod} was used for Rietveld refinements. A
Pseudo-Voigt profile shape function was used and full occupation
with no atomic disorder was considered for all positions.

The thermodynamic and transport physical properties were performed
by Physical Property Measurement System (PPMS) commercial device
(Quantum Design) and PPMS Dynacool (Quantum design) in the 2 – 300
K temperature range with applied magnetic field up to 9 T.
Specific heat was determined by means of the 2-$\tau$
relaxation method. Electrical resistivity and magnetoresistance were measured
using the 4-wire AC technique on the irregular samples shape in
relative units. Magnetic properties were performed by Magnetic
Property Measurement System (PPMS) (Quantum Design) in the
temperature range of 2 – 300 K under applied magnetic fields up to 9 T.
For temperature range below 1 K down to 80 mK, a $^3$He-$^4$He
dilution cryogen-free refrigerator TRITON 200 (Oxford Instruments, UK) with 8 T magnet was used.

\section{Results}

\subsection{Crystal structure of YbPd$_2$In}

\begin{figure}[tb]
  \centering
	\includegraphics[width=\linewidth]{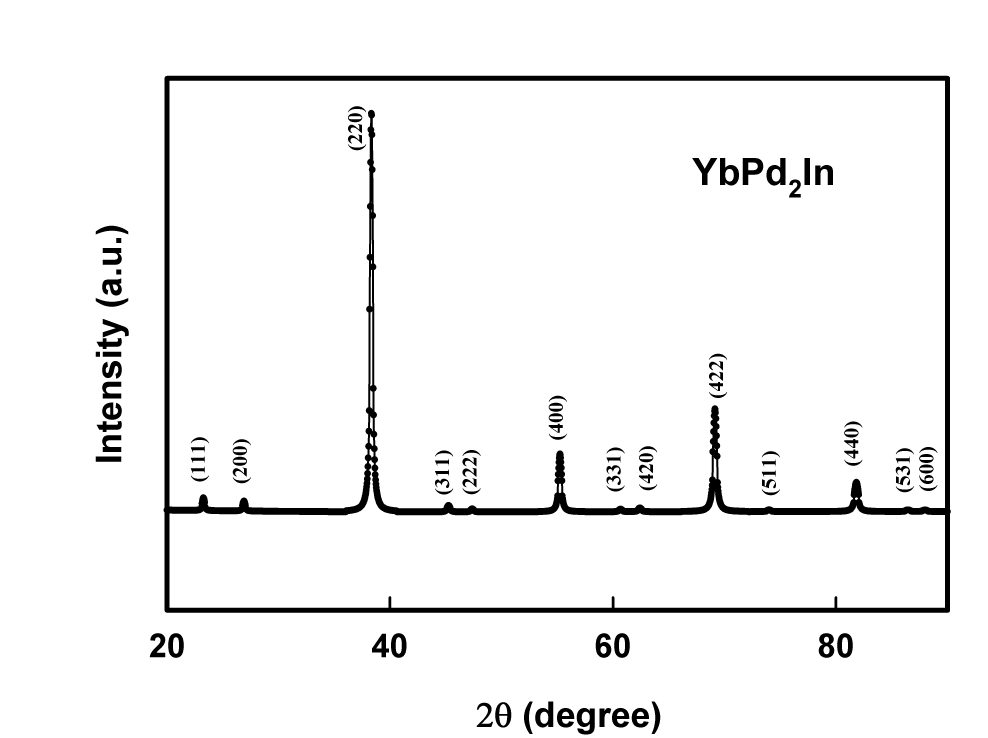}
  \caption{Powder x-ray diffraction patter for YbPd$_2$In at room temperature. All reflexes were indexed according to the cubic Heusleur phase structure type. Bragg peaks are indicated by the Miller indices.}
  \label{XRD}
\end{figure}

In the course of a systematic investigation of the ternary Yb-Pd-In system a
new ternary phase was found. A sample prepared on the
stoichiometry 1:2:1 from SEM/EPMA revealed to be practically
single phase. In fact, the XRD pattern of the compound (see
Fig.~\ref{XRD}) was successfully indexed by analogy with the
corresponding known cubic phase YbPd$_2$Sn, which crystallizes
with $\it{cF16}$ structure Cu$_2$MnAl-type (space group $\it{Fm\-\bar{3}m}$) with
lattice parameters a $=$ 6.661(5) \AA. Moreover, in agreement to
SEM results, in XRD pattern no peaks belonging to spurious phases
were found.

\subsection{Magnetic properties}

\begin{figure}
  \centering
  \includegraphics[width=\linewidth]{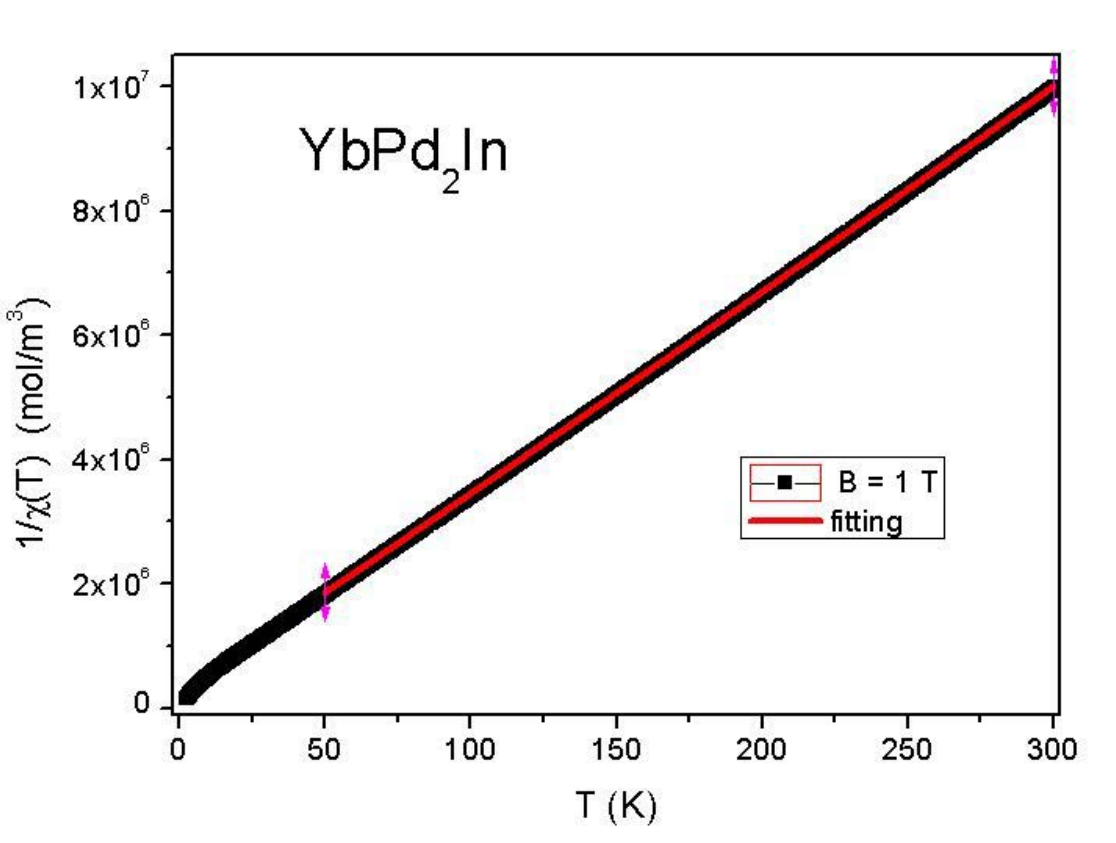}
  \caption{Temperature dependence of inverse susceptibility of YbPd$_2$In}
  \label{suscepti}
\end{figure}

The temperature dependence of $1/\chi(T)$ inverse magnetic
susceptibility for YbPd$_2$In is shown in Fig.
~\ref{suscepti}. The measurements were done in the
magnetic field of 1 T in the temperature range of 2–300 K. The
susceptibility data can be accounted for with a modified
Curie-Weiss (C-W) law given by the equation
\begin{equation} \label{Eq.1}
\chi(T)=\chi_0+\frac{c}{T-\Theta_P}
\end{equation}

From the high temperature fitting of the dependence $1/\chi(T)$
the value of the effective moment for YbPd$_2$In is $\mu_{eff} =
4.49$ $\mu_B$ which is close to the free Yb$^{3+}$ value (
$\mu_{eff} = 4.54$ $\mu_B$).  The paramagnetic Curie temperature
obtained from the fit is $\Theta_p = - 9$ K which is an indication
of antiferromagnetic exchange interactions. Notably, no
Pauli-like contribution $\chi_0$ can be extracted from the fit in
Fig. 2.

\subsection{Electrical resistivity}

\begin{figure}
  \centering
  \includegraphics[width=8cm]{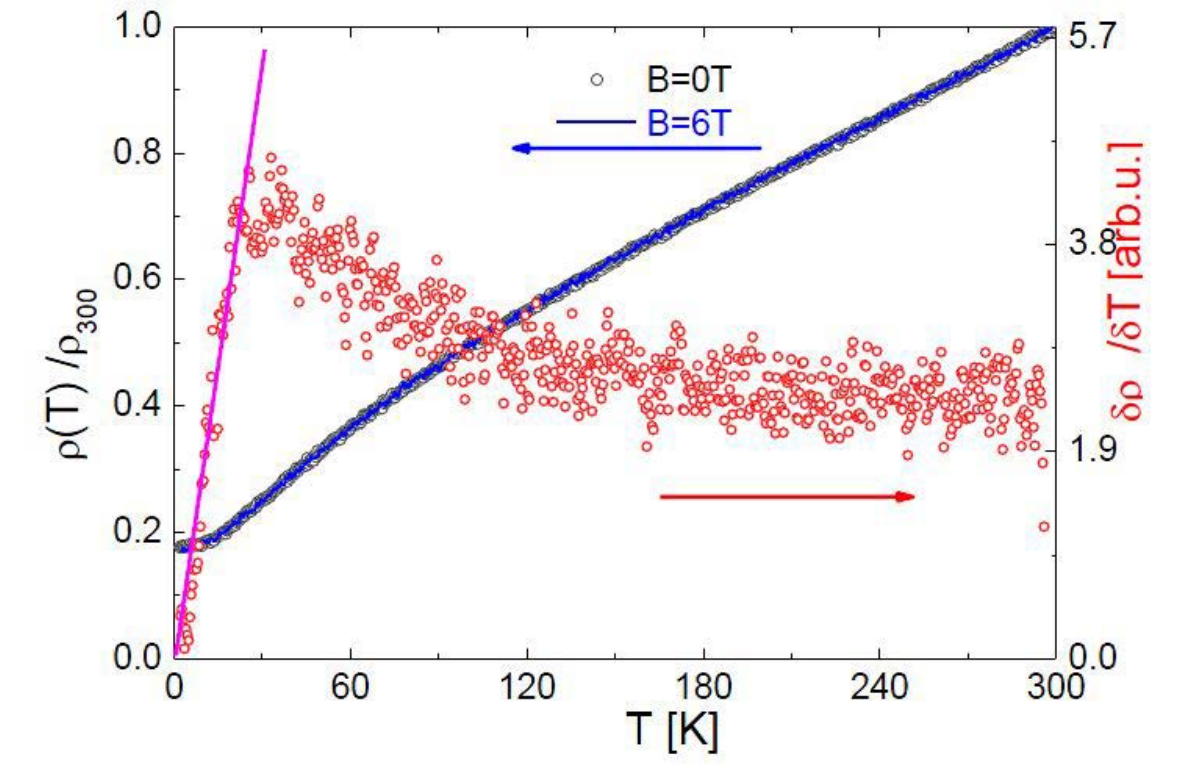}
  \caption{Left axis: temperature dependence of electrical
resistivity of YbPd$_2$In normalized at room (300K) temperature.
Right axis: electrical resistivity derivative showing a nearly
linear increase between zero and about 25K.}
  \label{resisti}
\end{figure}

The normalized temperature dependence of electrical resistivity
($\rho{(T)}$/$\rho_{300K}$) of YbPd$_2$In, measured at magnetic
fields of 0 T and 6 T,  is shown in Fig. ~\ref{resisti}. The
residual resistivity ratio RRR $\sim {5.5}$ indicates a good
quality of the polycrystalline sample, improving the RRR $\sim
{2.2}$ value of the YbPt$_2$Sn isotypic compound~\cite{14Gru} (the
comparison with YbPt$_2$In is meaningless because this compound
undergoes a CDW transition).

Focusing on  the the positive curvature of $\rho(T)$ at low
temperatures, the question arises about a coherent regime below
25 K. This feature is confirmed by the linear thermal variation of
the $\delta{\rho} / \delta{T}$ derivative, which corresponds to
the $\rho(T) = \rho_0 + A T^2$ dependence of a Fermi liquid (see
the straight line in Fig.~\ref{resisti}). On the other
hand, the broad negative curvature centered around 70 K could be
attributed to strongly hybridized CEF excited levels. However this
scenario is not reflected in the straight line thermal dependence
of $1/ \chi$ shown in Fig.~\ref{suscepti}. Moreover,
practically no magnetoresistance effect was detected. In fact,
applied fields of  B $=$ 3 T (not shown) and 6 T produce no
detectable effect on the the $\rho(T)$  dependence.

\subsection{Specific heat of YbPd$_2$In}

\begin{figure}
  \centering
  \includegraphics[width=\linewidth]{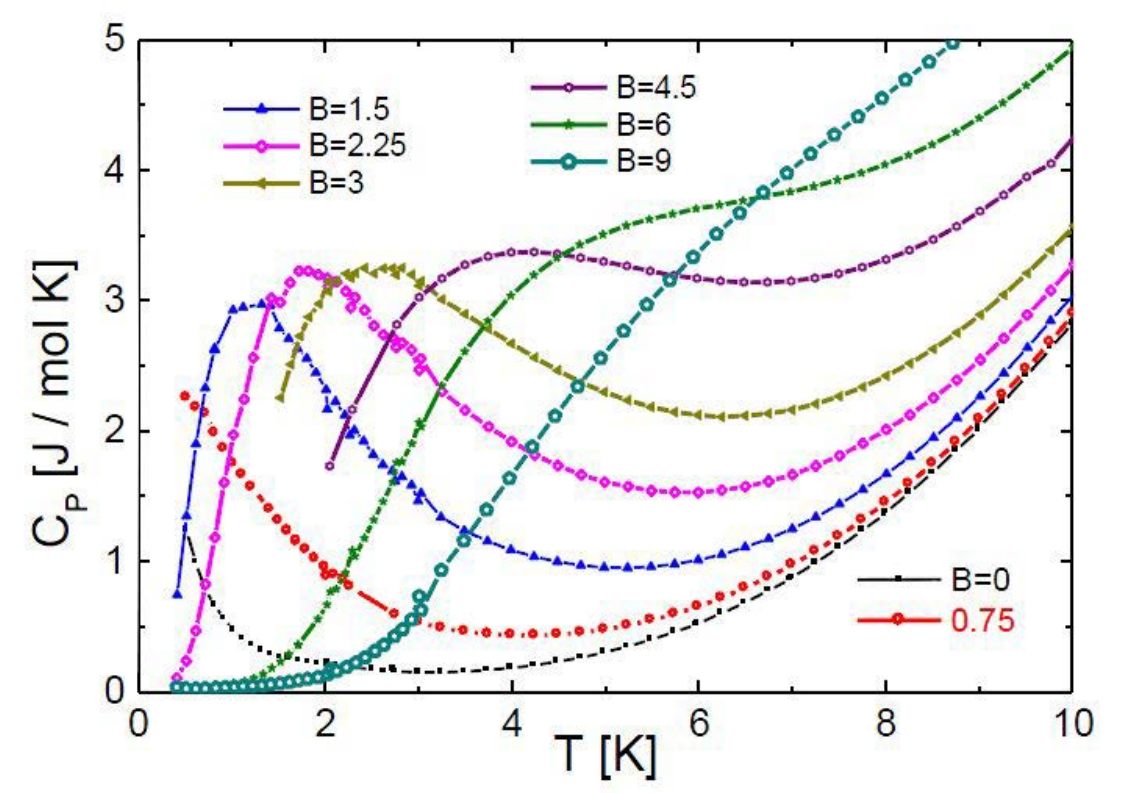}
  \caption{Thermal variation of the specific heat of YbPd$_2$In at zero and applied field up to B$=$9T.}
  \label{heatcapacity}
\end{figure}

In Fig.~\ref{heatcapacity} the specific heat $C_P(T)$
measurements performed at different magnetic fields (from 0 T to 9
T), within the $0.5<T<10$\,K range, are shown. For B $=$ 0 T a
minimum in the heat capacity at around 4 K can be seen from which
Cp(T) increases continuously by temperature decreasing. This
increase transforms into broad maximum centered at $T\approx
1.2$\,K for $B=1.5$\,T under magnetic field, which shifts to
higher temperature with increasing field intensity. These
anomalies can be qualitatively described as Schottky-type
anomalies which correspond to a two level system splitted by applied
field.

\begin{figure}
\centering
\includegraphics[width=7.5cm]{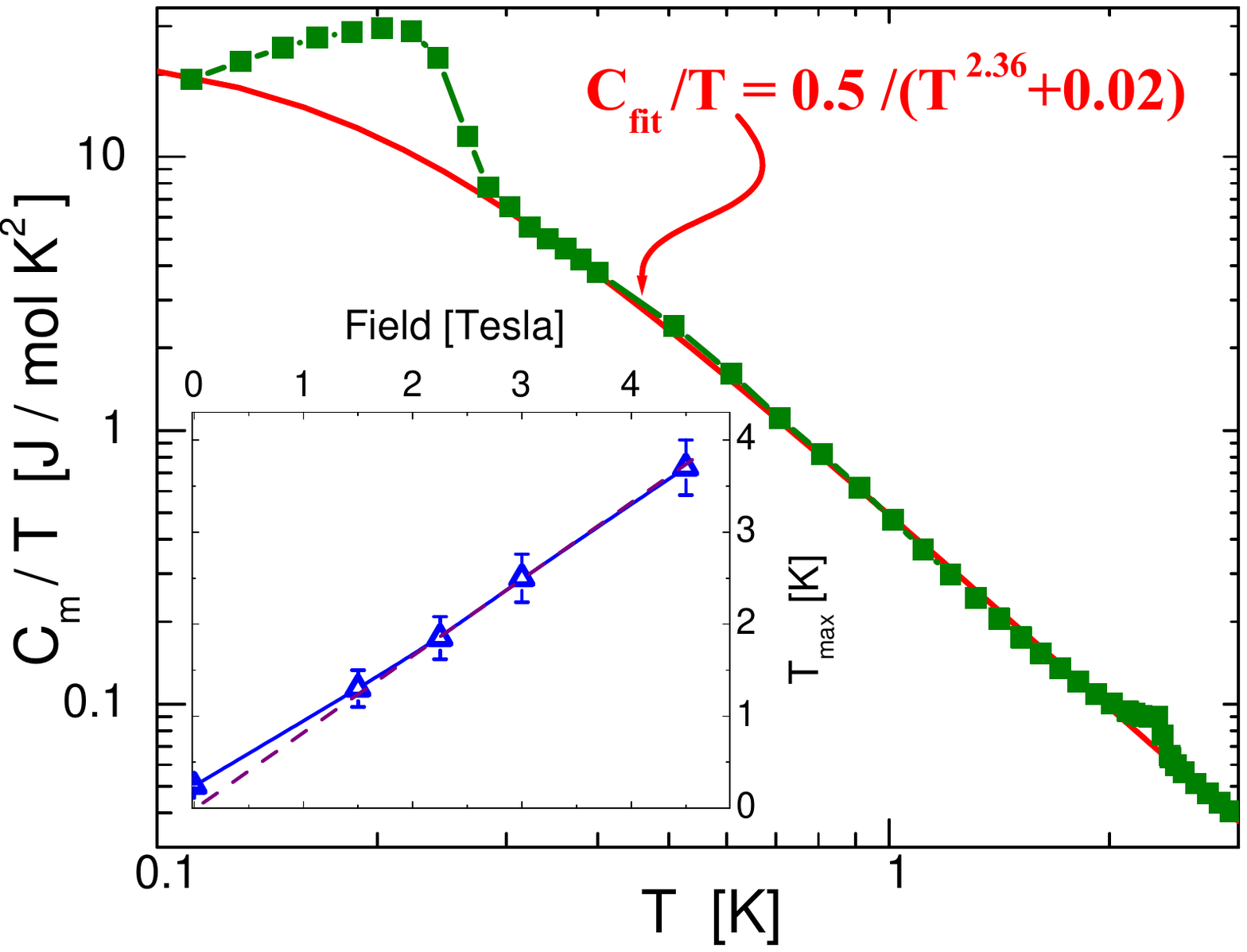}
\caption{$C_m/T$ vs T (at zero field) for YbPd$_2$In in a double
logarithmic representation. The continuous curve $C_{fit}/T$
represents a modified power law thermal dependence obtained within
the $0.28 <T< 3K$ range. Inset: $T_{max}$ with magnetic field, the
dashed line represents a linear extrapolation to zero.}
\label{Fig_lowTCp}
\end{figure}

Fig.~\ref{Fig_lowTCp} includes zero field $C_m(T)/T$ measurements
performed down to $T\approx 100$\,mK in a $log(C_m/T)$ versus
log(T) representation up to $T=3$\,K. The magnetic contribution to
specific heat $C_m/T$ was obtained after subtracting a phonon
contribution from the isotypic compound LuPt$_2$In
compound~\cite{15Jan} with a low temperature phonon coefficient
$\beta = 0.5$\,mJ/mol K$^3$. As it can be seen in the figure
$C_m/T$ decreases following a power law temperature dependence
down to $T\approx 280$\,mK where a jump in $C_m(T)$ occurs. At
$T\approx 210$\,mK $C_m/T$ reaches the highest registered value
for Yb-intermetallic compounds.

The small jump of $\Delta C_P= 53$\,mJ/molK observed at $T =
2.2$\,K is probably due to Yb oxide. It is worth 
noting that this effect is in coincidence with that reported for YbPd$_2$Sn ~\cite{85kie}
and associated to a superconductive transition. Notably, this
jump involves less than 1\% of the electronic degrees of freedom
and therefore it cannot be attributed to the Yb-4f electrons.

\section{Discussion}

%\subsection{Low temperature properties}

The effect of magnetic field on the specific heat upturn below
around 4\,K reveals the formation of a Schottky-like anomaly (see
Fig. ~\ref{Fig_heatcapacity}) with the temperature of its maximum
($T_{max}$) increasing proportionally to the field intensity:
$T_{max}/$K $\sim H/$Tesla (see the inset in
Fig.~\ref{Fig_lowTCp}), whereas the value of the maximum (at $C_P
\sim 3$\,J/mol K) barely increases.  Above the anomaly
$C_P(T>6K,B)$ grows significantly with field, suggesting an
increasing contribution of the excited CEF levels.

This Schottky-like anomaly in $C_P(T,B)$ indicates that the ground
state (GS) of this compound can be described as a two level system
with a splitting increasing proportionally to the field.
Within the error in determining the exact temperature on the broad
maxima, $T_{max}(B)$ extrapolates to $T_{max} = 0$ (see the dashed
line in the inset of Fig.~\ref{Fig_lowTCp}). However, since this
compounds shows a transition around $T^*=250$\,mK, it is evident
that the weak interaction responsible for the transition is
progressively overcome by the applied field (see the continuous
curve in the inset of Fig.~\ref{Fig_lowTCp}. The removal of the GS
degeneracy by magnetic field produces a significant shift of the
entropy to higher temperatures. This simple scenario may explain
why YbT$_2$X compounds (T$=$ Pd, Pt and X= In,Sn) are promising 
materials for adiabatic demagnetization in cooling
processes.

As expected, $C_m(T)$ at zero field is not described by a
Schottky-type anomaly. The analysis of the $C_m(T)/T$ dependence
above the transition (i.e. between $0.3 \leq T \leq 3$\,K) shows
that it can be properly fitted by a modified power law: $C_{fit}(T)/T
= 0.5/(T^{2.3}+0.02)$ as depicted in the double logarithmic
representation of Fig.~\ref{Fig_lowTCp}. This power law dependence
strongly resembles that of the isotypic compounds YbPt$_2$X (X =
In and Sn) ~\cite{14Gru}. In the case of these Pt homologues $C_m/T$
increases with similar trend down to $\approx 0.20$\,K where it
practically flattens in YbPt$_2$In and slightly turns down in
YbPt$_2$Sn.

%\subsection{Lack of magnetic order down to 500 mK}

Recent research has revealed a significant number of Yb-based intermetallic
compounds showing (or even not showing) magnetic order down to around 300\,mK,
most of them exhibiting a power law dependence above that
temperature ~\cite{14Gru, 14Cur, 15Gio, 17Ser}. Two main reasons, 
which can act also simultaneously, can be argued for this scenario: 
magnetic frustration and extremely weak magnetic interactions. 
Some crystalline structures, such as fcc cubic structures, provide the proper configuration for geometric frustration
~\cite{94Ram}.
%Nonetheless, these geometrical
%conditions do not apply to the compound under study because of its
%cubic Heusler type crystal structure.
The other inhibitor of long range magnetic order development
is the presence of very weak inter-site magnetic interactions as
proposed for the YbPt$_2$X homologues ~\cite{14Gru}. It is
well known that in intermetallic compounds the dominant mechanism
governing the magnetic exchange is the conduction
electrons-mediated RKKY interaction:

\begin{equation}   \label{eq:RKKY}
T_{RKKY} \sim J_{ex} * \delta(E_F) * f(1/d^3)
\end{equation}

where $J_{ex}$ is the coupling parameter, $\delta(E_F)$ the
density of the spin polarized band and $f(1/d^3)$ the envelopment
of the decreasing-oscillating function. Although the quite large
Yb-Yb spacing: $(d_{Yb-Yb})$ $= 4.7$~\AA~, may place YbPd$_2$In
within the weakly magnetic interacting systems, magnetic order
cannot be anyway excluded a priori for those interatomic spacings.

An empirical evaluation of the weight of the $\delta(E_F)$ factor
can be extracted from the Sommerfeld coefficient: $\gamma \sim
6$\,mJ/mol K$^2$, of LuPd$_2$In reported by in Ref.~\cite{15Jan}. This is
a small value considering that the formula unit contains four
atoms, whose respective individual $\gamma$ values add up more
than 30\,mJ/mol K$^2$. Thus, also the estimated $\delta(E_F)$ value for
YbPt$_2$X compounds place YbPd$_2$In within the weak
magnetic interacting systems.

The strength of the third factor: $J_{ex}$, can be evaluated from
the low $|\theta_P|_{T\to 0}$. From a low temperature fitting of 
$1/\chi(T)$ we obtain for YbPd$_2$In a value of $|\theta_P|_{T\to 0}$ = -0.98\,K.
This number together with the lack of any Kondo interaction
symptom also point to very weak $J_{ex}$ intensity. Although the
weakness of each one of these RKKY factors may guarantee the lack of
magnetic order, their multiplicative character easily explains the
small value of $T_{RKKY}$.

\section{Conclusions}

The newly synthesized cubic Heusler YbPd$_2$In compound was characterized in its
structural, magnetic, thermodynamic and transport properties. At
high temperatures the Yb atoms were found in their trivalent
$Yb^{3+}$ state. The absence of a significant Kondo effect is
demonstrated by the very low value of $\theta_P$ and the
continuous decrease of the electrical resistivity with
temperature.

At low temperatures this compound shows its relevant properties
with a very hight $C_m/T \approx 30$\,J/molK$^2$ value at
$T=200$\,mK, reached after a power law increase of $C_m(T)/T$
that collects nearly 1/2 of the
doublet GS entropy. Long range magnetic order is inhibited by a very weak
exchange interaction down to $\approx 300$\,mK. Under magnetic field, the $C_m(T,B)$
behavior is properly described by a simple two level scheme which
removes the GS degeneration by Zeeman splitting, whereas the weak
magnetic interaction is quenched. The resulting associated entropy is rapidly
shifted to higher temperature by the application of magnetic
field, making YbPd$_2$In a promising candidate as metallic refrigerant 
for adiabatic demagnetization cooling.

\section{Acknowledgments}
This work was supported by the projects VEGA 2/0032/16 and European Microkelvin Platform
APVV-17-0020.

\bibliography{references}
\bibliographystyle{apsrev4-1}

\end{document}